\documentstyle[psfig]{l-school}
\begin{document}
\markboth{F. Combes}{INTERACTIONS AND MERGERS}
\setcounter{part}{4}
%
\title{The Role of Interactions and Mergers}
\author{F. Combes}
\institute{Observatoire de Paris, 61 Av. de l'Observatoire, F-75014, Paris,
France}
\maketitle

\section{INTRODUCTION}

A long way has been run from the first views developped to explain
the formation of galaxies. In 1962, Eggen, Lynden-Bell \& Sandage 
designed the collapse scenario, where all galaxies are created with 
their morphological type, according to their angular momentum.
 Their potentials remained axisymmetric, so that no angular
momentum could be redistributed through gravity torques; the total
mass and gas content was already there at first collapse.
 For elliptical galaxies, the violent/single collapse picture
still remains in some modified form, although the most developped
and adopted scenario is through agglomeration of a large number
of clumps (e.g. van Albada 1982, Aguilar \& Merritt 1990), that 
produces de Vaucouleurs profiles in $r^{1/4}$. 
The merger picture (Toomre 1977, Schweizer 1990), where ellipticals
are formed by progressive interaction and coalescence of
many parent galaxies, is favored in hierarchical cosmogonies.

For spiral galaxies, the scenario involves now much more internal
dynamical evolution. Due to gas dissipation and cooling,
gravitational instabilities are continuously maintained in
spiral disks, and they drive evolution in much less than a 
Hubble time. Spiral galaxies are open systems, that accrete
mass regularly, and their morphological type evolves
along the Hubble sequence. Non-axisymmetric perturbations,
such as bars or spirals, produce gravity torques that drive
efficient radial mass flows; vertical resonances thicken disks
and form bulges, and the mass central concentration can destroy
bars. Accretion of small companions can also disperse bars
and enlarge the bulge. A major merger can destroy disks
entirely and form an elliptical.

The first role of galaxy interactions is to trigger internal
evolution, that we will consider now, in the next section. 
Specific aspects of galaxy interactions and mergers
will then be detailed in section \ref{envir}.

\section{INTERNAL PROCESSES}
\label{dyn}

\subsection{Gravitational Viscosity}

During their life, galaxies accrete and concentrate mass;
this requires that angular momentum 
is redistributed, and essentially is tranfered in the outer
parts. For interstellar gas, viscous torques could be thought of 
to get rid of the angular
momentum. However, normal viscosity is not efficient, due to the
very low density of the gas. Even with 
macroturbulent viscosity, the time-scales are
longer than the Hubble time at large radii, and could 
be effective only inside the central 1kpc
(Lynden-Bell \& Pringle 1974).

The other efficient way to transfer angular momentum
is through gravity torques, due to non-axisymmetric
perturbations, or density waves. In the stellar component,
trailing spirals transport momentum outwards 
(Lynden-Bell \& Kalnajs 1972), and if the spiral is open
(i.e. the pitch angle large enough), there exists a
phase-shift between the stellar density and the potential
of the spiral, such that a torque is exerted on the 
particles, and momentum is redistributed (Zhang 1996).
This process is even more important in the gas component,
since it is more responsible to gravitational instabilities,
due to efficient cooling. The coupled star-gas components
are more unstable than each separately (e.g. Jog 1992, Romeo 1992).
The gravity torques due to instabilities
have been called "gravitational viscosity" (Lin \& Pringle 1987a).

Gravitational instabilities are suppressed at small scales
through the local velocity dispersion $c$, and at large
scale by rotation. The corresponding limiting scales
are the Jeans scale for a 2D disk $\lambda_J \sim c^2/(G \mu)$,
and $\lambda_c \sim G \mu/\kappa^2$, where $\mu$ is the 
disk surface density, and $\kappa$ the epicyclic frequency.
Scales between $\lambda_J$ and $\lambda_c$ are unstable,
unless $c$ is larger than $\pi G \mu/\kappa$, or the Toomre
$Q = { {c \kappa} \over {\pi G \mu} }$ is larger than 1.
 If the disk is cold at the beginning (general case for the gas),
instabilities set in, which heat the disk until $Q \sim 1$,
and those instabilities provide the necessary angular momentum
tranfer (or viscosity) to concentrate the mass. Since
the size of the region over which angular momentum is tranferred
is $\sim \lambda_c$, and the time-scale is a rotation period,
$2 \pi / \Omega$,
the effective kinematic viscosity is $\nu \sim \lambda_c^2  \Omega$,
and the typical viscous time 
$\tau_{\nu} \sim R^2 \Omega^3 / (G^2 \mu^2)$.

\subsection{Exponential Disks}

A galactic disk, submitted to gravitational instabilities
can then be compared to an accretion disk.
In the theory of viscous accretion disks,
the main effect of viscosity is to move angular-momentum
outward and mass inward. In the particular hypothesis of
equal time-scales between viscosity and star formation,
i.e. $\tau_{\nu} \sim \tau_*$, it can be shown that the
final surface density of the stellar disk is exponential
(Lin \& Pringle 1987b). 
An exponential distribution of metallicity
can also be derived in these circumstances
Tsujimoto et al. (1995).

The viscous and star-formation time-scales are
of the same order of magnitude, since the two processes depend
exactly on the same physical mechanism, i.e. gravitational
instabilities. As shown empirically by Kennicutt (1989),
the Toomre parameter $Q$ appears to control star-formation
in spiral disks. Therefore, if the regulating instabilities
have time to develop, one can expect that 
 $\tau_{\nu} \sim \tau_*$, as required for exponential
light and metallicity distribution.

\subsection{Bars and Nuclear Bars}

A bar is the most efficient non-axisymmetric perturbation
able to drive radial gas flows. But a single dynamical
structure can act only over a limited range of radii,
between the two most characteristic Lindblad resonances; to have
a continuous action over a large radial range, and drive
the gas to the nucleus, multiple-scale perturbations
are required, such as bars within bars (e.g. Shlosman et al 1989).
Bars are frequently triggered by galaxy interactions,
and once the matter has been driven towards the center,
the inner disk becomes in turn unstable to bar formation.
Nested bars are often observed in spiral disks (e.g. Jarvis 
et al 1988, Wozniak et al 1995), and even more easily in 
the near-IR band (Shaw et al 1993, Friedli et al 1996,
Jungwiert et al 1997). One of their main characteristics
is that the nuclear bars are inscribed inside the nuclear 
ring of the primary bar (corresponding to its inner Lindblad
resonance), and that they can have any orientation with
respect to the primary bar. 

Numerical simulations have described the numerical processus
leading to their formation (Friedli \& Martinet 1993, Combes 1994). 
In concentrating the mass towards the center, the first bar
modifies the inner rotation curve, and the precessing rate
($\Omega - \kappa/2$)
of the $m=2$ elliptical orbits in the center is elevated
to large values. This strong differential precessing rate
prevents the self-gravity from matching all precessing rates
in the center, and decoupling occurs: two bars rotating
at two different pattern speeds grow. In the simulations,
the two bars have a resonance in common, most often the inner
Lindblad resonance of the primary bar is the corotation of
the secondary one. It is probable that the two bars exchange
energy at this common resonance, and that is the reason of
their growth (Tagger et al 1987).

\section{GALAXY INTERACTIONS}
\label{envir}

Tidal interactions are the main external trigger of dynamical
processes that induce galaxy evolution. They produce
non-axisymmetrical perturbations, that concentrate mass
and trigger star-formation. Starbursts are observed
when a huge gas mass is concentrated in the center of the 
interacting/merging system.
CO emission suggest that the molecular gas represents a significant
part of the dynamical mass there (Scoville et al 1991). 
 Non nuclear starbursts are very rare (cf Stanford et al 1990, 
Yun et al 1994). 
 To trigger such starbursts, gas must be brought towards the center
in a time-scale short enough with respect to the feedback time-scale
of star-formation (a few 10$^7$ yr), that will 
blow the gas back outwards (e.g. Larson 1987).
It is through numerical simulations that insight has been gained in those
systems, although many uncertainties remain about the physics of
gas and star formation (see the review by Barnes \& Hernquist 1992).

\subsection{Numerical Codes and Star Formation }

It is necessary to take into account all components self-consistently,
stars, gas and dark matter. The latter participates actively in 
dynamical friction and receives the extra angular momentum,
allowing the visible galaxies to merge.

Gas dissipation is a key factor in the formation of density
waves and non-axisymmetric structures,
although viscous torques are negligible versus the gravity torques
(Combes et al. 1990). 
We dont know precisely the actual viscosity of the ISM,
but given its very complex multi-phase and  small-scale structure, 
it is not relevant to model it in any accurate manner. Any large-scale
hydrodynamical simulation can reproduce the main
characteristics of gas flow in galaxies, provided that viscous torques are 
negligible. Two families of gas modelisations are currently used,
one based on a continuous diffuse fluid, essentially governed by 
pressure forces. Such modelisations include artificial viscosity to
spread shock waves over a few resolution cells. The physics of the gas
is assumed to be isothermal at 10$^4$K (case of SPH or finite difference codes).
The other modelisation used in galaxy hydrodynamics is the sticky particles
approach, where an ensemble of gas clouds move in ballistic orbits and
collide, without extra pressure and viscosity terms. This modelisation 
represents more closely the fragmented structure of the molecular component.

 As for star-formation, the numerical codes are very schematical,
since the detailed processes (SFR, IMF) are still unknown.
 Since stars are observed to be formed inside
giant molecular clouds in our Galaxy, the latter being the result
of agglomerations of smaller entities, one process could be to 
relate star formation to cloud-cloud collisions, in the sticky particles
modelisation (Noguchi \& Ishibashi 1986).  
Another more widely used is to adopt a
Schmidt law for the SFR, i.e. the rate is proportional to a power $n$ of the
gas volumic density, $n$ being between 1 and 2 (Mihos et al 1992).
 In both cases, it was shown that interacting galaxies were the
site of strong starbursts, that could be explained both by the
orbit crowding in density waves triggered by the tidal interactions,
and by the gas inflow and central concentrations, accumulating the
gas in small and very dense regions. This depends of course on the non-
linearity of the Schmidt law, and SF-efficiency strongly depends on
the power $n$ (see e.g. Mihos et al 1992). Mihos \& Hernquist (1994)
use a hybrid-particles techniques, within SPH, to describe the effects
of gas depletion and formation of a young star population. The SPH/young
star particles are converted from gaseous to collisionless form, as soon
as their gas mass fraction drops below 5\%. Although this method has
many computational advantages (number of particles fixed), it does
not decouple the behaviour of young stars and gas, especially in the shocks.
 This also inhibits contagious star formation, or any feedback
mechanisms (Struck-Marcell \& Scalo 1987, Parravano 1996).

\subsection{ Major and Minor Mergers}

 During the interaction of galaxies of comparable mass,
strong non-axisymmetric forces are exerted on the interstellar gas. 
But contrary to what could be expected, the main torques responsible for the 
gas inflow are not directly due to the companion, but to
internal processes triggered by it. The tidal perturbations destabilise
the primary disk, and the non-axisymmetric structures generated in the 
primary disk (bars, spirals) are responsible for the torques. 
 The self-gravity of the primary disk, and its consequent
gravitational instabilities, play the fundamental role.
The gas fuel is provided by the primary disk itself, and not
by the companion.
  This is why the first parameter determining the characteristics
of the merger event is the initial mass distribution in the two
interacting galaxies (Mihos \& Hernquist 1996). The mass
ratio between the bulge and the disk is a more fundamental
parameter than the geometry of the encounter.

 The central bulge stabilises the disk with respect to external
perturbations. If the bulge is sufficiently massive, the apparition
of a strong bar is delayed until the final merging stages, and so
is the gas inflow, and the consequent star-formation activity.
 But the starburst can then be stronger. When the primary disk is
of very late type, without any bulge, the gravitational instability
settles in as soon as the beginning of the interaction, there
is then a continuous activity during the interaction, but 
at the end the starburst is then less violent, since most
of the gas has been progressively consumed before.

In the simulations, about 75\% of the gas is consumed during the
merger, whatever the internal structure of galaxies, or the 
geometry of the encounter (Mihos \& Hernquist 1996). This is
the most uncertain parameter, however, since the physics
of the gas is only schematically reproduced, with too much 
viscosity.
 What is the fate of the rest of the gas? In general
long tidal tails are entrained in the outer parts,
especially in neutral hydrogen, since this is the 
most abundant component in external parts of galaxies.
But most of the material of the tails is still bound
to the system, and will rain down progressively onto
the merger remnant (Hibbard 1995). 

\smallskip

 During less equal galaxy interactions, 
where the mass ratio between the two colliding 
galaxies is at least 3, the same features can be noticed:
the first relevant parameter is the mass concentration in the 
galaxies before the interaction (Hernquist \& Mihos 1995).
The torques responsible for the gas mass inflow are exerted by 
the non-axisymmetric (essentially $m=2$)
potential developped in the disk. 

\subsection {Tidal Tails and Dark Matter}

 The extent of tidal tails can help to constrain the amount of dark 
matter around galaxies. Simulations have shown that, as the dark-to-luminous
mass ratio increases, the length of the tails and the mass involved in them
is considerably reduced (Dubinski et al 1996). Larger masses imply higher
speed encounters, that will detune resonances between the angular frequency
of the orbital and internal motions, required to form tails. The deeper
potential wells to climb induce shorter tails. Simulations are compatible
with observations for dark-to-luminous mass ratios between 0 and 8, but 
not higher. This corresponds to the dark matter detected from HI rotation
curves, but rules out more massive determinations, such as from timing
argument for the Milky Way (10$^{12}$ M$_{\odot}$ from the orbit of our 
companion M31).

\subsection {Interactions and Disks}

Galaxy interactions can easily thicken or even destroy a 
stellar disk (e.g. Gunn 1987). The fragility of stellar disks with
respect to thickening has been used by Toth \& Ostriker (1992)
to constrain the frequency of merging and the value of the cosmological
parameter $\Omega$. They claim for instance that the Milky Way disk have
accreted less than 4\% of its mass within the last 5 10$^9$ yrs. 
Numerical simulations have tried to quantify the thickening effect
(Quinn et al 1993, Walker et al 1996). They show that the stellar disk
thickening can be large and sudden, but it is strongly moderated by 
gas hydrodynamics and star-formation processes, since the thin disk can 
be reformed continuously through gas infall. Galaxies presently interacting
have their ratio $h/z_0$ of the radial disk scalelength $h$ to the
scaleheight $z_0$ 1.5 to 2 times lower than normal
(Bottema 1993; Reshetnikov \& Combes 1997). 
 However, since galaxies have experienced many interactions in the past,
including the presently isolated galaxies, all these perturbations, 
thickening of the planes and radial stripping, must be transient,
and disappear after an interaction time-scale, i.e. one Gyr. Present
galaxies are thought to be the result of merging of smaller units, according
to theories of bottom-up galaxy formation; a typical galaxy has accreted
most of its mass, and the existence of shells and ripples
attests of the frequency of interactions (Schweizer \& Seitzer 1992).
 This implies that the global thickness of galaxy planes can recover
their small values after galaxy interactions. Or in other words,
the disk of present day spirals has been essentially assembled at low 
redshift (Mo et al 1998).

\subsection{ Ring Galaxies }

Ring galaxies, like the Cartwheel, are believed to
be formed during a head-on encounter between a disk galaxy and a companion
crossing its plane (e.g. Lynds \& Toomre 1976, Theys \& Spiegel 1976).
They constitute an ideal laboratory to understand the dynamics and physics
of the gas in galaxy collisions, since a single impulse is given
by the companion, and the geometry of the phenomenon is very simple, 
(see e.g. Appleton \& Struck-Marcell 1996). 

Ring galaxies are not among the most violent starburst
($>$ 100 M$_\odot$/yr), but they reveal about 10 times the 
star formation activity of normal galaxies. 
They are among the rare objects
where the starburst is non-nuclear.
 In the Cartwheel, the H$\alpha$ comes exclusively from the
outer ring (Higdon 1995), and 80\% of it comes from just
one quadrant, quite asymmetrically. The recent star formation rate
(H$\alpha$) is about 10 times that over the last 15 Gyrs (B-band),
and the consumption time-scale of the HI gas (although 
abundant 1.3 10$^{10}$ M$_\odot$) is only 430 Myrs, of the
same order of the ring expansion time-scale. 

There are many unsolved problems in ring galaxies, and in
the Cartwheel in particular:
 the region of the maximum HI density in the ring does not
coincide with the region of maximum star formation (Higdon 1996);
but of course we still do not know the H$_2$ distribution
(cf Horellou et al 1995). Also, there is no gas in the 
spokes, nor in the center (Higdon 1996), while N-body simulations of 
the encounter predict there a gas concentration
(e.g. Hernquist \& Weil 1993).

\subsection {Groups and Clusters}

Although the galaxy volumic density is larger in clusters,
the relative velocity also, and
tidal interactions might be less effective than in the field.
Distorted galaxies with asymmetries, tails and plumes are often observed 
in the stellar component (which cannot be due to ram-pressure),
but starbursts are less frequent, since they require a 
large abundance of gas in the outer
parts of galaxies, available for gas inflow, and this gas is 
removed by tidal stripping. It is heated and forms then the coronal
gas observed in X-ray in compact groups and clusters. In the case
of strong gas deficiency, the interaction frequency can even
be anti-correlated with star-formation rate.

This is true even in compact groups, where the relative velocity
is not too high.
 There is some (controversed) evidence of excess far-infrared emission
(Zepf 1993, Sulentic et al 1993) indicating more star-formation than 
isolated galaxies (especially the 60/100$\mu$ colors). There is also
excess radio emission (Menon 1995) and HI deficiency (Williams et al 1991), 
which confirm their peculiarity with respect
to field galaxies. A recent ROSAT survey of 22 HCGs (Ponman et al 1996) 
detected the hot coronal gas through X-ray radiation in 75\% of them.
But there is no evidence of higher molecular gas 
content in most HCG galaxies (Boselli et al 1996, Leon et al 1998).
  Therefore, although there exist some obvious signs of interaction,
galaxy collisions have limited efficiency (there is no merging), and limited
star-formation triggering.

The effect of gas stripping is even more important in rich clusters
( e.g. Cayatte et al 1990).
In Virgo, there is evidence that the global star formation
rates have been reduced (Kennicutt 1983). 
The accumulation of frequent high-speed close encounters, 
dubbed "harassment" by Moore et al (1996), has peculiar
consequences, quite different from the normal galaxy-galaxy
binary interactions and merging. Simulations have shown
that these close encounters, although they are not resonant
with the internal motions, can produce the tidal damage
observed in most perturbed galaxies in clusters (e.g. Combes
et al 1988). One issue is to distinguish between the influence
of the global tidal action of the cluster, and the two-body
encounters between individual galaxies. The former global effect
is expected to be significant only in the central part, while
the best cases of tidal deformations are observed on the periphery.

 Harassment is so efficient that it is possible to account for the
rapid galaxy evolution in rich clusters (Moore et al 1996): within 4-5
Gyr, the morphology of galaxies in clusters can be transformed from
late-type spirals to early-type, lenticulars and ellipticals. Only
4-5 high-speed collisions are necessary for this evolution.

\section{EVOLUTION WITH REDSHIFT}

\subsection {Observations of High-z Galaxies}

\begin{figure}[t]
\psfig{figure=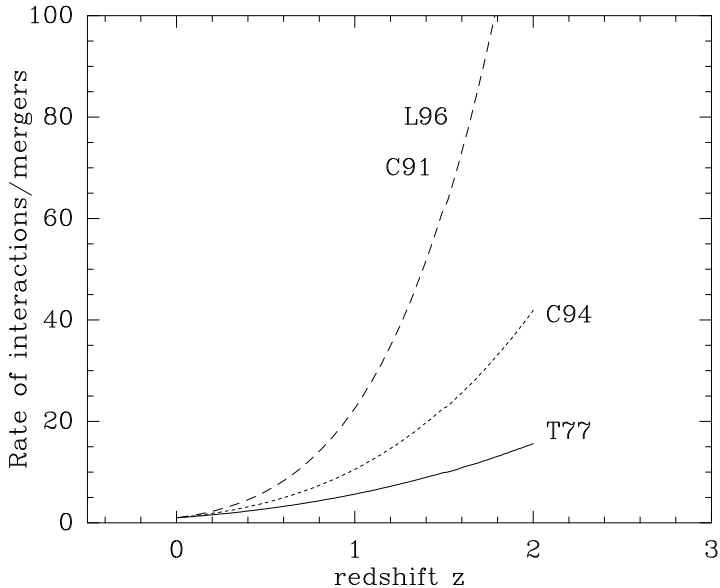,bbllx=0mm,bblly=3cm,bburx=18cm,bbury=18cm,width=10cm}
\caption{ Evolution of the rate of interaction and mergers as a function
of redshift, from various estimations. Solid line: estimation from the
present number or merger remnants in the NGC catalog, and frequency of 
apparition of binary periods $P$ in P$^{-5/3}$ from Toomre (1977, T97); no evolution
is postulated here. Dashed line: from the frequency of collisional ring galaxies
in deep HST fields (Lavery et al 1996, L96) and various other objects 
(pairs, Butcher-Oemler effet, IRAS faint sources, QSOs, Carlberg 1991, C91). 
Dotted line: from the number of interacting pairs (Carlberg et al 1994, C94).
 The observed power laws (for $z < 1$) have been extrapolated to $z = 2$. }
\label{fig1}
\end{figure}

The advent of the Hubble Space Telescope has allowed,
through high resolution imaging, to study galaxy 
morphology at intermediate redshift (0.1 $\le$ z $\le$ 1.0).
A striking feature is that high-z galaxies are often seen with
tidal features, distorted morphologies indicative of interactions and
mergers. This might be surprising since tidal arms and distortions
are generally weak features, of relatively short life-time, and their 
detectability is reduced at high redshift (Mihos 1995). 
The debris and signatures of mergers become 
invisible after 1 Gyr for $z=0.4$ and after 0.2 Gyr only at $z=1$. 
 The fact that tidal distortions are so omnipresent in high-z
galaxies, in spite of this bias,  is a clue to the huge increase 
of interactions and mergers with $z$.

In clusters, the relative importance of "nature or nurture" in
accounting for the large percentage of elliptical galaxies,
is still a debated question: are elliptical galaxies formed
at once from rapid collapse in rich environment that are
to become clusters, or are they the result of late merging of disk 
galaxies, when the cluster virializes?
High-resolution images of clusters at high $z$ can bring some light
to this problem. Dressler et al (1994) have imaged a cluster at $z=0.4$, and
suggest that the excess blue galaxies seen in distant clusters are predominantly
normal late-type spirals, undergoing tidal interactions or mergers.
The Butcher-Oemler effet, associated with enhanced star-forming activity
appears now convincingly to be due to enhanced interactions, arising
from hierarchical merging. Clusters at $z=0.4$, i.e. 3-4 Gyr
ago, appear to possess a much larger fraction of disk-dominated galaxies
than present day clusters (Couch et al 1994).

In the frame of hierarchical cosmological scenarios, it is expected
that the number of galaxies in a comoving volume increases with
redshift, while their mass decreases. Some trends in that sense
have been revealed by HST at very high redshift (z $\approx$ 2-3).
A population of faint, small, compact objects has been identified
to the sub-galactic clumps, or building blocks of present day galaxies
through multiple merging processes (Pascarelle et al 1996, Steidel et al 
1996). However, caution must be used to interprete high redshift data,
since fading of low surface brightness features (as $(1+z)^{-4}$), and
k-correction effects transform galaxies in later types and more irregular
objects, even without evolution (Giavalisco et al 1996). For instance,
van den Bergh et al (1996) have claimed
that barred and grand design galaxies appear under-represented 
in distant deep field samples, with respect to early-types galaxies.
But this could be only the bias against detecting disks at high $z$.
At very high redshifts, only the high surface density objects,
such as bulges and ellipticals, can be detected. 

Galaxy interactions were undoubtedly more frequent in the
past, and many groups have tried to quantify the effect.
Already Toomre in 1977 has estimated the number of mergers
from their observed frequency at $z=0$ just taking into
account the probability of excentricities of binary orbits.
Statistics of close galaxy pairs from faint-galaxy redshift surveys
have shown that the merging rate increases as a power law with
redshift, as $(1+z)^{m}$ with $m=4\pm1.5$ (e.g. Yee \& Ellingson 1995). Lavery
et al (1996) claim that ring galaxies are also rapidly evolving,
with $m=4-5$, although statistics are still insufficient.
Many other surveys, including IRAS faint sources, 
or quasars, have also revealed a high power-law (see fig \ref{fig1}).

\subsection{Different Evolution Time-scales}

We already see at $z=0$ a wide range of galaxy properties, 
that appear to correspond to various stages of evolution.
The main parameter determining the Hubble sequence is the bulge-to-disk 
luminosity ratio, which increases from late to early type spiral
galaxies. In other words, the mass concentration, and therefore the
evolution stage, increases from Sc to S0. From recent near-IR surveys,
de Jong (1996) and Courteau et al (1996) claim that the bulge mass
is even a better criterion than the bulge-to-disk ratio to define the sequence.
The total mass increases from Sc to S0, and might be
the essential parameter (e.g. Gavazzi et al 1996). 
Late-type galaxies and
irregulars possess a much larger gas fraction than early-types,
which is also a clue of their low evolution. Galaxy
evolution is indeed driven both by dynamical processes, that concentrate
the mass (see section \ref{dyn}), and  star formation that
consumes the available cold gas. 

Also in the recent years, we have realized the importance of
Low Surface Brightness galaxies (LSB), which appear unevolved
systems, with large gas fraction, and low mass concentration
(Bothun et al 1997). Compared to High Surface Brightness galaxies
(HSB), they are more dominated by dark matter, even within 
the optical disks, as are dwarf irregulars (de Blok \& McGaugh 1997).
 The clue to their long evolution time-scales might be their
poor environment, and low interaction frequency (Bothun et al 1997).

N-body simulations have shown that
bars and density waves through their gravity torques
can concentrate galaxy masses on a time-scale much shorter than the 
Hubble time. Since most spiral galaxies are observed with these features,
this implies rapid evolution along the Hubble sequence.
But along the sequence, galaxies gain visible mass, and their 
dark-to-luminous mass ratio decreases: this suggests that some of their
dark matter has been transformed in visible stars (Pfenniger et al 1994,
Pfenniger \& Combes 1994).


\begin{thebibliography}{100}  
\bibitem{} Aguilar L.A., Merritt D. \review ApJ, 354, 1990, 33
\bibitem{} Appleton P.N., Struck-Marcell C. \review Fund. of Cosmic Phys.,
    16, 1996, 111
\bibitem{} Barnes J.E., Hernquist L. \review ARAA, 30, 1992, 705
\bibitem{} Boselli A., Mendes de Oliveira C., Balkowski C. et al \review 
        A\&A, 314, 1996, 738
\bibitem{} Bothun G., Impey C., McGaugh S. \review PASP, 109, 1997, 745
\bibitem{} Bottema R. \review A\&A, 275, 1993, 16
\bibitem{} Carlberg R.G. \review ApJ, 375, 1991, 429
\bibitem{} Carlberg R.G., Pritchet C.J., Infante L. \review ApJ, 435, 1994, 540
\bibitem{} Cayatte V., Balkowski C., van Gorkom J.H., Kotanyi C. \review  
   AJ, 100, 1990, 604
\bibitem{} Combes F., Dupraz C., Casoli F., Pagani L. \review 
              A\&A, 203, 1988, L9
\bibitem{} Combes F., Dupraz C., Gerin M. \book in Dynamics and Interactions
            of Galaxies, ed. R. Wielen, Heidelberg: Springer Verlag, 1990,  205
\bibitem{} Combes F. \book in Mass-transfer induced activity in galaxies, 
                     ed. I. Shlosman, Cambridge Univ. Press, 1994, 170
\bibitem{} Couch W.J., Ellis R.S., Sharples R.M., Smail I.
              \review ApJ, 430, 1994, 121
\bibitem{} Courteau S., de Jong R.S., Broeils A.H. \review ApJ, 457, 1996, L73 
\bibitem{} de Blok W.J.G., McGaugh S.S. \review MNRAS, 290, 1997, 533
\bibitem{} de Jong R.S. \review A\&A, 313, 1996, 45
\bibitem{} Dressler A., Oemler A., Butcher H.R., Gunn J.E. \review ApJ, 
             430, 1994, 107
\bibitem{} Dubinski J., Mihos J.C., Hernquist L. \review ApJ, 462, 1996, 576
\bibitem{} Friedli D., Martinet L. \review A\&A, 277, 1993, 27
\bibitem{} Friedli D., Wozniak H., Rieke M. et al \review A\&AS, 118,
 1996, 461
\bibitem{} Gavazzi G., Pierini D., Boselli A. \review A\&A, 312, 1996, 397
\bibitem{} Giavalisco M., Livio M., Bohlin R.C. et al
          \review AJ, 112, 1996, 369
\bibitem{} Gunn J.E. \book in Nearly Normal Galaxies, ed. S.M. Faber, 
             Springer: New York, 1987, 459
\bibitem{} Hernquist, L., Weil M.L \review MNRAS, 261, 1993, 804.
\bibitem{} Hernquist, L., Mihos, J.C. \review ApJ, 448, 1995, 41.
\bibitem{} Hibbard J.E. (1995) PhD thesis, Columbia University
\bibitem{} Higdon J.L. \review ApJ, 455, 1995, 524
\bibitem{} Higdon J.L. \review ApJ, 467, 1996, 241
\bibitem{} Horellou C., Casoli F., Combes F., Dupraz C. \review A\&A, 298, 1995, 743
\bibitem{} Jarvis B., Dubath P., Martinet L., Bacon R. \review A\&AS, 74, 1988, 513
\bibitem{} Jog C. \review ApJ, 390, 1992, 378
\bibitem{} Jungwiert B., Combes F., Axon D. \review A\&AS, 125, 1997, 479  
\bibitem{} Kennicutt R.C. \review AJ, 88, 1983, 483
\bibitem{} Kennicutt R.C. \review ApJ, 344, 1989, 685
\bibitem{} Larson R.B. \book in Starbursts and galaxy evolution,
      ed. T.X. Thuan T. Montmerle and J. T. T. Van, Ed. Fronti\`eres,  1987, 467
\bibitem{} Lavery R., Seitzer P., Walker A.R. et al
              \review ApJ, 467, 1996, L1
\bibitem{} Leon S., Combes F., Menon T.K. \review A\&A, 330, 1998, 37
\bibitem{} Lin D.N.C, Pringle J.E. \review MNRAS, 225, 1987a, 607  
\bibitem{} Lin, D.N.C., Pringle, J.E. \review ApJ, 320, 1987b, L87  
\bibitem{} Lynden-Bell, D., Kalnajs, A.J. \review MNRAS, 157, 1972, 1
\bibitem{} Lynden-Bell, D., Pringle J.E. \review MNRAS, 168, 1974, 603
\bibitem{} Lynds B., Toomre A. \review ApJ, 209, 1976, 382
\bibitem{} Menon T.K. \review MNRAS, 274, 1995, 845
\bibitem{} Mihos J.C. \review ApJ, 438, 1995, L75
\bibitem{} Mihos J.C., Hernquist L. \review ApJ, 437, 1994, 611
\bibitem{} Mihos, J.C., Hernquist, L. \review ApJ, 464, 1996, 641.
\bibitem{} Mihos J.C., Richstone D.O., Bothun G.D. \review ApJ, 400, 1992, 153
\bibitem{} Mo H.J., Mao S., White S.D.M. \review MNRAS, 295, 1998, 319
\bibitem{} Moore B., Katz N., Lake G. et al \review Nature,
         379, 1996, 613
\bibitem{} Noguchi M., Ishibashi S. \review MNRAS, 219, 1986, 305
\bibitem{} Parravano A. \review ApJ, 462, 1996, 594
\bibitem{} Pascarelle S.M., Windhorst R.A., Keel W.C., Odewahn S.C. 
              \review Nature, 383, 1996, 45
\bibitem{} Pfenniger D., Combes F., Martinet L. \review A\&A, 285, 1994, 79
\bibitem{} Pfenniger D., Combes F. \review A\&A, 285, 1994, 94
\bibitem{} Ponman T., Bourner P., Ebeling H., Bohringer H. \review MNRAS,
283, 1996, 690
\bibitem{} Quinn P.J., Hernquist L., Fullagar D.P. \review ApJ, 403, 1993, 74
\bibitem{} Reshetnikov V., Combes F. \review A\&A, 324, 1997, 80
\bibitem{} Romeo A.B. \review MNRAS, 256, 1992, 307
\bibitem{} Schweizer F. \book in Dynamics and Interactions
            of Galaxies, ed. R. Wielen, Heidelberg: Springer Verlag, 1990,  60
\bibitem{} Schweizer F., Seitzer P. \review AJ, 104, 1992, 1039 
\bibitem{} Scoville, N.Z., Sargent, A.I., Sanders, D.B., Soifer, B.T. 
    \review    ApJ, 366, 1991, L5
\bibitem{}  Shaw, M.A., Combes, F., Axon, D.J., Wright, G.S. \review A\&A,
273, 1993, 31
\bibitem{} Shlosman I., Frank J., Begelman M. \review Nature, 338, 1989, 45
\bibitem{} Stanford S.A., Sargent A.I, Sanders D.B., Scoville N.Z.
        \review ApJ, 349, 1990, 492
\bibitem{} Steidel  C.C., Giavalisco M., Dickinson M., Adelberger K.L.
            \review AJ, 112, 1996, 352
\bibitem{} Struck-Marcell C., Scalo J.M. \review ApJS, 64, 1987, 39
\bibitem{} Sulentic J.W., de Mello Raba\c ca D. \review ApJ, 410, 1993, 520
\bibitem{}  Tagger, M., Sygnet, J.F., Athanassoula, E., Pellat, R. \review ApJ,
 318, 1987, L43
\bibitem{} Theys J.C., Spiegel E.A. \review ApJ, 208, 1976, 650
\bibitem{} Toomre A. \book in The Evolution of Galaxies and Stellar Populations,
ed. B.M. Tinsley \& R.B. Larson, New Haven: Yale Univ. Obs., 1977, 401
\bibitem{} Toth G., Ostriker J.P. \review ApJ, 389, 1992, 5
\bibitem{} Tsujimoto T., Yoshii Y., Nomoto K., \& Shigeyama T. \review 
A\&A, 302, 1995, 704
\bibitem{} van Albada T.S. \review MNRAS, 201, 1982, 939
\bibitem{} van den Bergh S., Abraham R.G., Ellis R.S. et al 
                  \review AJ, 112, 1996, 359
\bibitem{} Walker, I.R., Mihos, J.C., Hernquist, L. \review ApJ, 460, 1996, 121
\bibitem{} Williams B.A., MacMahon P., van Gorkom J. \review AJ, 101, 1991, 1957
\bibitem{} Wozniak H., Friedli D., Martinet L. et al \review
A\&AS, 111, 1995, 115
\bibitem{} Yee H.K.C., Ellingson E. \review ApJ, 445, 1995, 37
\bibitem{} Yun M.S., Scoville N.Z., Knop R.A. \review ApJ, 430, 1994, L109
\bibitem{} Zepf S.E. \review ApJ, 407, 1993, 448
\bibitem{} Zhang X. \review  ApJ,  457, 1996, 125
\end{thebibliography}
\end{document}